\documentclass[aps, prd, 10pt, notitlepage, superscriptaddress, nofootinbib,numbers,showpacs]{revtex4-1}

\usepackage[utf8]{inputenc}
\usepackage[T1]{fontenc}
\usepackage{anyfontsize}
\usepackage{amsmath}
\usepackage{amssymb}
\usepackage{amsfonts}
\usepackage{mathrsfs}

\usepackage{microtype}

\usepackage{graphicx}
\usepackage{epsfig}

\usepackage[dvipsnames]{xcolor}
\usepackage{hyperref}
\hypersetup{
    colorlinks=true,
    citecolor=Purple,
    linkcolor=Purple,
    urlcolor=Purple,
    linktocpage=true,
    breaklinks=true
}

\usepackage[capitalize]{cleveref}
\begin{document}
\title{
LQG inspired spacetimes as solutions of the Einstein equations}

\author{Marcos V. de S. Silva}
	\email{marcosvinicius@fisica.ufc.br}
	\affiliation{Departamento de F\'isica, Universidade Federal do Cear\'a, Caixa Postal 6030, Campus do Pici, 60455-760 Fortaleza, Cear\'a, Brazil.}
\author{Carlos F. S. Pereira}
	\email{carlosfisica32@gmail.com}
	\affiliation{Departamento de F\'isica, Universidade Federal do Esp\'irito Santo, Avenida Fernando Ferrari, 514, Goiabeiras, 29060-900, Vit\'oria, ES, Brazil.}
\author{G. Alencar}
\email{geova@fisica.ufc.br}
\affiliation{Departamento de F\'isica, Universidade Federal do Cear\'a, Caixa Postal 6030, Campus do Pici, 60455-760 Fortaleza, Cear\'a, Brazil.}
\author{Celio R. Muniz}
\email{celio.muniz@uece.br}
\affiliation{Universidade Estadual do Cear\'a (UECE), Faculdade de Educa\c{c}\~ao, Ci\^encias e Letras de Iguatu, Av. D\'ario Rabelo s/n, Iguatu - CE, 63.500-00 - Brazil.}

\date{today; \LaTeX-ed \today}
\begin{abstract}
Black bounces are compact objects with a wormhole structure hidden behind an event horizon. This type of metric can be obtained through general relativity by considering the presence of exotic matter. Such spacetimes can also arise within the framework of effective theories inspired by loop quantum gravity. In this work, we verify the possibility of obtaining black bounce models inspired by loop quantum gravity as solutions of general relativity. For this, we examine which sources can generate these solutions and the consequences of using these types of sources. We find that the sources can be expressed as a combination of a phantom scalar field and nonlinear electrodynamics. Once we obtain the sources in terms of fields, we analyze the energy conditions for each field separately to verify which of the fields is responsible for the violation of the energy conditions.
\end{abstract}
\maketitle
\def\HMS{{\scriptscriptstyle{HMS}}}
\section{Introduction}
\label{S:intro}
The theory of general relativity (GR) proposed by Albert Einstein in 1915 represents a revolution in modern thinking regarding the concept of gravity, which is no longer interpreted as a force and is now interpreted as a geometric effect reactive to matter fields. This is one of the most successful physical theories in terms of experimental and observational verifications \cite{INTRO1, INTRO2, INTRO3, INTRO4, INTRO5, INTRO6, INTRO7, INTRO8, INTRO9, INTRO10}. Therefore, its acceptance as a starting point for the investigation of gravitational systems is unanimous among experts in the field. Several spherically symmetric models presenting geodesic singularity problems were proposed within this important gravitational theory, describing the so-called black hole (BH) solutions \cite{Schwarzschild:1916uq}. In this context, the first and simplest proposed solution was that of Schwarzschild, who described a vacuum solution for a spherically symmetric and static BH. Subsequently, several other singular models emerged for spherically symmetric and static systems and spherically symmetric and stationary systems. Although they commonly exhibit singularities in their interior, the main characteristic of a BH is the presence of an event horizon, a no-return surface that causally separates distinct regions of spacetime.

In addition to these systems that presented problems regarding the presence of singularity, another class of models, now without the presence of singularity, was initially proposed by James Bardeen around 1968 to represent the geometry of a regular black hole (RBH) \cite{INTRo11}. Later, around 2000, Beato and Garcia obtained the source of matter for this prototype, which was nonlinear electrodynamics (NED), in which the regularization parameter begins to be interpreted as the charge of the magnetic monopole \cite{INTRO12}. Still in the context of these RBH solutions, several other models were proposed, such as the electrically charged case analyzed by Rodrigues and Silva \cite{INTRO13} among others \cite{INTRO14, INTRO15, INTRO16, INTRO17, INTRO18, INTRO19, INTRO20, INTRO21, INTRO22, INTRO23, INTRO24, INTRO25}. 

Still in the scenario of regular solutions, recently Simpson and Visser proposed a new class of compact objects that describe some variations of wormhole (WH) solutions depending on the relationship between the model parameters, which are the mass \textbf{m} and the radius of the throat \textbf{a} \cite{Simpson:2018tsi}. Such objects were called black bounces (BB) and can present some possible configurations according to the relationship between the model parameters. Thus, if $a=2m$ we have the formation of a unidirectional WH, if $a>2m$ we have the case of a bidirectional WH, and for $a<2m$ we have the formation of a WH throat hidden by two symmetric event horizons. The source of matter associated with this regularized gravitational system that does not satisfy vacuum conditions was later found to be a combination of NED and phantom scalar fields \cite{Canate:2022gpy,Bronnikov:2021uta,Rodrigues:2023vtm}. Thus, after this class of BB solutions emerged, several other gravitational scenarios were explored to verify possible new BB solutions. Scenarios include k-essence theories \cite{CF1,CF2,CF3}, modified gravity \cite{GRAVMOD1,GRAVMOD2,GRAVMOD3,GRAVMOD4}, conformal gravity \cite{GRAVCONFORM}, teleparallel gravity \cite{GRAVTELL}, black strings \cite{GEO1,GEO2,GEO3,Geo4}, spherically symmetric and stationary backgrounds \cite{ROTA1,ROTA2,GEO2}, among others \cite{OUTROS1,OUTROS2,OUTROS3,outros4,outros5}. These BB models are being analyzed through theoretical and observational tests such as for gravitational lensing effects in the weak and strong field regime \cite{BBOBS1,BBOBS2,BBOBS3,BBOBS4,BBOBS5,BBOBS6,BBOBS7,BBOBS8,BBOBS9,BBOBS10,Heidari:2024bkm}, quasi-normal modes, shadows, and optical appearance and scattering \cite{BBOBS11,BBOBS12,BBOBS13,BBOBS14,BBOBS15}.

From this perspective, Loop Quantum Gravity (LQG) emerges as a compelling framework for describing gravitational systems beyond GR, particularly in regimes where quantum effects become significant. It provides tools to investigate compact objects, both with and without geodesic singularities, providing insight into the quantum structure of spacetime. A key feature is the polymer representation \cite{Modesto:2008im,Laddha:2010wp}, which enables the mapping of singular spacetimes to fully regular geometries through effective quantum corrections.

Although LQG does not yet constitute a complete quantum theory of spacetime, particularly near classical singularities, its applications have been predominantly concentrated in semiclassical regimes. Substantial progress has been made in modeling singularity resolution and building effective quantum-corrected geometries \cite{LQG1,LQG2,LQG3,LQG4,LQG5,Lewandowski:2022zce}, including configurations involving WH structures \cite{Muniz:2024jzg,deSSilva:2024gdc}. In this context,~\cite{LQG6} proposed a procedure to introduce LQG-inspired corrections into already regular spacetimes. This method, based on a coordinate transformation reminiscent of the approach by Simpson and Visser \cite{Simpson:2018tsi}, yields regular geometries that incorporate nontrivial quantum gravitational effects, as exemplified in \cite{Muniz:2024wiv}. Notably, in the latter reference, the resulting RBHs can be interpreted as structures supported by sources modeled through NED within the framework of GR. These developments broaden the applicability of LQG-based models, providing new pathways for constructing physically realistic, non-singular solutions within a semiclassical gravitational framework.

The effective matter sector reproducing a given LQG-inspired metric is not unique. This metric itself is already provided as a smooth effective geometry that encodes the coarse-grained influence of the underlying discrete quantum structure of spacetime \cite{BenAchour:2018khr}. Since these geometries describe a continuum regime in which the microscopic LQG degrees of freedom are no longer resolved, and in which a matter sector is required at sufficiently large scales \cite{Kelly:2020uwj}, classical GR provides a natural framework in which the metric can be reinterpreted as inducing an effective stress–energy tensor. Our aim is therefore to provide a consistent and minimal classical realization that embeds the LQG-inspired geometry within GR. This reconstruction is useful because it ensures that the spacetime satisfies Einstein’s equations, permits the discussion of physical properties in a familiar macroscopic setting, and yields a consistent framework for perturbative analyses. Although different effective decompositions may exist, as long as they reproduce the same background geometry, the geometric sector of the perturbations -- typically the physically relevant one -- remains determined by the metric itself. We adopt thus a GR non-vacuum reconstruction (phantom scalar + NED) for the LQG-inspired geometry because it allows a direct perturbative treatment with standard GR tools. The linearized system reduces to second-order wave equations with well-posed boundary conditions, so QNMs, possible echoes, and stability windows can be computed with standard time and frequency domain numerical solvers \cite{Duran-Cabaces:2025sly, Franzin:2023slm}. In contrast, modified gravity models often lead to higher-order or coupled equations, making a well-posed formulation harder and requiring custom numerics.

Thus, we consider the Simpson-Visser spacetime as a starting point to test the formalism required to find the source fields of BHs with LQG corrections as solutions of GR. We constructed sources for electrically and magnetically charged solutions, which were a combination of NED and phantom scalar fields, and then the energy conditions were constructed and analyzed. In our approach, LQG-inspired spacetimes are not treated as fundamentally quantum, but rather as effective geometries that can be consistently described in terms of classical GR. Once such metrics are specified, the corresponding stress-energy tensor can be reconstructed, revealing that the necessary sources may be effectively realized by a phantom scalar field in combination with NED. These classical fields thus provide an effective representation of the energy–momentum content needed to sustain quantum-inspired corrections, as also confirmed by the Simpson-Visser case taken here as a benchmark example. Understanding these sources can be fundamental when studying the properties of the solutions. By interpreting the sources, we can investigate in greater depth the perturbations of the metric and of the source fields. This can reveal more information about the stability of these spacetimes than merely analyzing stability with test fields \cite{Franzin:2023slm}.

The present study first establishes in section \ref{S:Fieldeq} the static spherically symmetric line element inspired by LQG for a gravitational system where the scalar field is minimally coupled to gravity and an electrodynamic function. The equations of motion and general electrodynamic functions are derived for the magnetic and electric cases. In sections \ref{S:SV}, \ref{S:HCS}, and \ref{S:POLY} the consistency of the method used was tested by applying it to electrically and magnetically charged systems, and then all quantities of interest were derived, all of them having LQG corrections. In section \ref{S:EnerCond}, the energy conditions were analyzed for the Schwarzschild BH with Holonomy correction and the polymerized BH. Finally, section \ref{S:conclusion} summarizes the main conclusions regarding the validity of this method, as well as future perspectives.


 We adopt the metric signature $(+,-,-,-)$.
 We shall work in geometrodynamics units where $8 \pi G=\hbar=c=1$. 

\section{Field source to a general spacetime}
\label{S:Fieldeq}
As the main point of this work is to find sources that can generate LQG inspired spacetimes in GR, we will consider the line element that can be written as
\begin{equation}
    ds^2=g(r)dt^2-f(r)^{-1}dr^2-r^2\left(d\theta^2+\sin^2\theta d\varphi^2\right).\label{lineel}
\end{equation}
The components of the Einstein tensor, $G_{\mu\nu}$, for this type of line element are written as
\begin{eqnarray}
    {G^{0}}_0&=&-\frac{f'}{r}-\frac{f}{r^2}+\frac{1}{r^2},\label{G00}\\
    {G^{1}}_1&=&-\frac{f g'}{r g}-\frac{f}{r^2}+\frac{1}{r^2},\label{G11}\\
    {G^{2}}_2&=&{G^{3}}_3=-\frac{f' g'}{4 g}-\frac{f'}{2 r}-\frac{f g''}{2 g}+\frac{f g'^2}{4
   g^2}-\frac{f g'}{2 r g}\label{G22}.
\end{eqnarray}
In the context of GR, such a general solution could not be described only by a scalar field or NED. This can be seen through the symmetries of the stress-energy tensor for each field source. In the case of the scalar field, there is the symmetry ${T^0}_0={T^2}_2$ and in the electromagnetic case, we have the symmetry ${T^0}_0={T^1}_1$. 

In general, it is always possible to find metrics as solutions of Einstein equations for a given energy density and generic pressures. However, what we seek here are more concrete examples of sources, rather than generic ones, that can generate such solutions. This type of problem also arises in the case of BB spacetimes, such as the Simpson-Visser model. As is usually done in the Simpson-Visser model, we will check whether the coupling of gravitational theory together with NED and a phantom scalar field is capable of generating this spacetime.

The action that describes this type of theory is given by
\begin{equation}
    S=\int d^4x\sqrt{-g}\left[R-2\epsilon g^{\mu\nu}\partial_\mu\phi\partial_\nu\phi+2V\left(\phi\right)+L(F)\right],\label{action}
\end{equation}
with $R$ being the Ricci scalar, $g$ is the determinant of the metric tensor $g_{\mu\nu}$, $V\left(\phi\right)$ is the potential related to the scalar field $\phi$ and the electromagnetic Lagrangian $L(F)$ is an arbitrary function of the electromagnetic scalar $F=F^{\mu\nu}F_{\mu\nu}$. According to the notation we are following in this work, if $\epsilon=1$ we have a usual scalar field and if $\epsilon=-1$ we have a phantom scalar field. 
 
From the action \eqref{action}, we find the following field equations
\begin{eqnarray}
 &&   \nabla_\mu \left[L_F F^{\mu\nu}\right]=\frac{1}{\sqrt{-g}}\partial_\mu \left[\sqrt{-g}L_F F^{\mu\nu}\right]=0,\label{eqmax}\\
 &&   2\epsilon \nabla_\mu \nabla^\mu\phi=-\frac{dV(\phi)}{d\phi},\label{eqscal}\\
   && {R^\mu}_{\nu}-\frac{1}{2}\delta^\mu_{\nu}R={T^\mu}_{\nu}[\phi]+{T^\mu}_{\nu}[F],\label{eqGR}
\end{eqnarray}
where $L_F=\partial L/\partial F$ and $R_{\mu\nu}$ is the Ricci tensor. The stress-energy tensors ${T^\mu}_{\nu}[\phi]$ and ${T^\mu}_{\nu}[F]$ are given by
\begin{eqnarray}
    {T^\mu}_{\nu}[\phi]&=&2\epsilon \partial^\mu\phi\partial_\nu\phi-\delta^\mu_{\nu}\left(\epsilon g^{\alpha\beta}\partial_\alpha \phi \partial_\beta \phi-V(\phi)\right),\\
    {T^\mu}_{\nu}[F]&=&\frac{1}{2}\delta^\mu_{\nu}L(F)-2L_F {F}^{\mu\alpha} F_{\nu \alpha}.
\end{eqnarray}

Depending on the type of source chosen, the Maxwell-Faraday tensor may have a nonzero component $F_{23}$, for the magnetic case, or $F_{01}$, for the electric case. 

\subsection{Magnetic case}
For magnetic charged solutions, the only nonzero component of the Maxwell-Faraday tensor, $F_{\mu\nu}$, is
\begin{equation}
    F_{23}=q_m \sin \theta,\label{f23}
\end{equation}
and the electromagnetic scalar is
\begin{equation}
    F(r)=\frac{2q_m^2}{r^4},\label{EMScalar}
\end{equation}
where $q_m$ is the magnetic charge.

From \eqref{f23} and \eqref{EMScalar}, the modified Maxwell equations, equations \eqref{eqmax}, are identically satisfied. The remaining equations of motion are
\begin{eqnarray}
 -\frac{f'}{r}-\frac{f}{r^2}+\frac{1}{r^2}=V+\frac{L}{2}+\epsilon  f \phi
   '^2,\label{eqGR1}\\
  f \left(\epsilon  \phi '^2-\frac{\frac{r
   g'}{g}+1}{r^2}\right)+\frac{1}{r^2}=\frac{L}{2}+V,\label{eqGR2}\\
 \frac{-r^3 g \left(r f' g'+2 f \left(r
   g''+g'\right)\right)-2 r^3 g^2 f'+r^4 f
   g'^2}{4 r^4 g^2}=\frac{L}{2}-\frac{2 q_m^2
   L_F}{r^4}+V+\epsilon  f \phi '^2,\label{eqGR3}\\
   -\epsilon  f' \phi '+\epsilon  f
   \left(\left(-\frac{g'}{g}-\frac{4}{r}\right) \phi '-2 \phi
   ''\right)=-\frac{V'}{\phi '}.\label{eqphi}
\end{eqnarray}

From equations \eqref{eqGR2} and \eqref{eqGR3}, we can isolate the electromagnetic quantities and write them as
\begin{eqnarray}
   L(r)&=&2 \left(f \left(-\frac{g'}{r g}-\frac{1}{r^2}+\epsilon  \phi
   '^2\right)+\frac{1}{r^2}-V\right),\label{L_gen_magnetic}\\
   L_F(r)&=&\frac{r^2 \left(r g \left(r f' g'-2 f \left(g'-r
   g''\right)\right)+2 g^2 \left(r f'+f \left(4 r^2
   \epsilon  \phi '^2-2\right)+2\right)-r^2 f g'^2\right)}{8 q_m^2
   g^2}.\label{LF_gen_magnetic}
\end{eqnarray}

For quantities associated with the scalar field, it is not yet possible to write the functions $V(r)$ and $\phi(r)$ in closed form. What we can obtain from equations \eqref{eqGR1} and \eqref{eqphi} are the derivatives of these functions, which are
\begin{eqnarray}
      V'(r)&=&\epsilon  \phi ' \left(f' \phi '+f
   \left(\left(\frac{g'}{g}+\frac{4}{r}\right) \phi '+2 \phi
   ''\right)\right).\label{Vgeneral}\\
  \phi'(r)&=&\frac{i \sqrt{g f'-f g'}}{\sqrt{2r\epsilon
   fg}},\label{scalarderi}
\end{eqnarray}
with $i$ being the unit imaginary number.

Therefore, we need to specify the spacetime model we are considering and finally obtain the functions $\phi(r)$ and $V(r)$.


\subsection{Electric case}
If we consider an electric charged source, only the component $F_{01}=-F_{10}$ of the Maxwell-Faraday tensor is nonzero. From the modified Maxwell equations, \eqref{eqmax}, we find
\begin{equation}
    F^{10}=\frac{q_e}{L_F r^2}\sqrt{\frac{f}{g}}, \quad \mbox{and} \quad F=-\frac{2 q_e^2}{r^4 L_F^2},\label{F_eletric}
\end{equation}
where $q_e$ is the electric charge.

For an electrically charged source, the Einstein equations undergo modifications only in the electromagnetic contribution, and thus are written as:
\begin{eqnarray}
   -\frac{f'}{r}-\frac{f}{r^2}+\frac{1}{r^2}=\frac{L}{2}+\frac{2 q_e^2}{r^4 L_F}+\epsilon  f \phi
   '^2+V,\\
   -\frac{f \left(\frac{r g'}{g}+1\right)}{r^2}+\frac{1}{r^2}=\frac{L}{2}+\frac{2 q_e^2}{r^4
   L_F}-\epsilon 
   f \phi '^2+V,\\
   \frac{g \left(r f' g'+2 f \left(r
   g''+g'\right)\right)+2 g^2 f'-r
   f g'^2}{4 r g^2}=-\epsilon  f \phi
   '^2-\frac{L}{2}-V.
\end{eqnarray}

As the part of the equations of motion related to the scalar field remains the same, only the electromagnetic functions undergo modifications and are given by:
\begin{eqnarray}
    L(r)=\frac{-g \left(r f' g'+2 f \left(r
   g''+g'\right)\right)-2 g^2 f'+r
   f g'^2}{2 r g^2}+2 f \phi '^2-2 V,\label{L_gen_electric}\\
   L_F(r)=\frac{8 q_e^2 g^2}{r^2 \left(r g \left(r f'
   g'-2 f \left(g'-r g''\right)\right)+2
   g^2 \left(r f'+f \left(-4 r^2 \phi
   '^2-2\right)+2\right)-r^2 f g'^2\right)}\label{LF_gen_electric},
\end{eqnarray}
which are different from equations \eqref{L_gen_magnetic} and \eqref{LF_gen_magnetic}.

Although obtained independently through the Einstein equations, the electromagnetic functions $L$ and $L_F$, for both the electric and magnetic cases, are related to each other, since $L_F$ is the derivative of the electromagnetic Lagrangian with respect to the electromagnetic invariant, and must obey the consistency relation:
\begin{equation}
    L_F=\frac{\partial L}{\partial F}= \frac{\partial L}{\partial r}\left(\frac{\partial F}{\partial r}\right)^{-1} \rightarrow L_F-\frac{\partial L}{\partial r}\left(\frac{\partial F}{\partial r}\right)^{-1}=0\label{EMbound}.
\end{equation}

As we shall see later, for electrically charged solutions, it becomes more complicated, but not always impossible, to analytically construct the Lagrangian $L(F)$. In these cases, it is simpler to work with the auxiliary field $P_{\mu\nu} = L_F F_{\mu\nu}$. Associated with this field, we have the invariant $P$, which is given by:
\begin{equation}
    P=P^{\mu\nu}P_{\mu\nu}=L_F^2 F=-\frac{2q_e^2}{r^4}.\label{P_electric}
\end{equation}
In general, the above function is much easier to invert than the scalar $F$, the equation \eqref{F_eletric}.

If an electrodynamics theory tends to Maxwell in the weak-field limit, we have $L_F \to 1$, and thus
\begin{equation}
    L(F) \approx F \approx P.
\end{equation}
Thus, in the weak-field limit, the two representations are identical. However, for BHs, the associated electrodynamics often does not reduce to Maxwell's theory in the limit $F \to 0$.


\section{Simpson--Visser Solution}\label{S:SV}
In order to test the consistency of the method, we will first consider the Simpson-Visser BB solution, which has a bounce structure similar to LQG-inspired solutions. Usually, the line element that describes this spacetime is written as \cite{Simpson:2018tsi}
\begin{equation}
    ds^2=\left(1-\frac{2m}{\sqrt{a^2+\rho^2}}\right)dt^2-\left(1-\frac{2m}{\sqrt{a^2+\rho^2}}\right)^{-1}d\rho^2-\left(\rho^2+a^2\right)\left(d\theta^2+\sin^2\theta d\varphi^2\right).\label{lineSV}
\end{equation}
The constant $a$ is the regularization parameter of this model and represents the radius of the throat.

Despite presenting itself differently from the equation \eqref{lineel}, through a coordinate transformation, we can rewrite this line element as the equation \eqref{lineel} with the functions $f(r)$ and $g(r)$ given by \cite{Canate:2022gpy}
\begin{equation}
   g(r)=\left(1-\frac{2m}{r}\right) , \qquad f(r)=\left(1-\frac{2m}{r}\right)\left(1-\frac{a^2}{r^2}\right).\label{coef_SV}
\end{equation}

Using the functions from \eqref{coef_SV} in \eqref{L_gen_magnetic}-\eqref{scalarderi} and considering that we have a phantom scalar field, $\epsilon=-1$, we find

\begin{eqnarray}
    \phi(r)&=&\tan ^{-1}\left(\frac{\sqrt{r^2-a^2}}{a}\right),\qquad    V(r)=\frac{4 a^2 m}{5 r^5},\label{V-SV}\\
    L(r)&=&\frac{12 a^2 m}{5 r^5}, \quad
    L_F(r)=\frac{3 a^2 m}{2 q_m^2 r}, \quad F(r)=\frac{2q_m^2}{r^4}.\label{LF-SV}
\end{eqnarray}
With appropriate adaptations of the coordinates and notation, the results are compatible with those obtained in previous works \cite{Canate:2022gpy,Bronnikov:2021uta,Rodrigues:2023vtm}. The electromagnetic quantities satisfy the relation \eqref{EMbound}.

\begin{figure}
    \centering
    \includegraphics[width=0.5\linewidth]{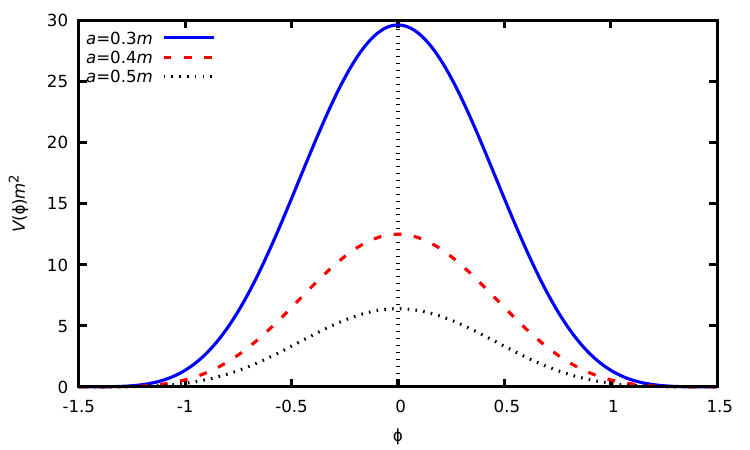}
    \caption{Behavior o the scalar field potential to the Simpson-Visser solution as a function of $\phi$ considering different values of $a$.}
    \label{fig:Vphi_SV}
\end{figure}

Using the expressions for $F(r)$ and $\phi(r)$, we can write the electromagnetic Lagrangian and the potential associated with the scalar field as:
\begin{equation}
    L(F)=\frac{3\ 2^{3/4} a^2 F^{5/4} m}{5 q_m^{5/2}} , \qquad V(\phi)=\frac{4 m}{5 a^3 \sec (\phi )^{5}}.
\end{equation}
Expanding $V(\phi)$ to $\phi \approx 0$, we find
\begin{equation}
    V(\phi) \approx \frac{4 m}{5 a^3}-\frac{2 m \phi ^2}{a^3}.
\end{equation}
Thus, we see that at zero the potential tends to a constant, and the first correction term is proportional to $\phi^2$, similar to the correction of a massive scalar field potential. In Fig. \ref{fig:Vphi_SV}, we see the behavior of the potential, which resembles a potential barrier that grows as the value of $a$ decreases.

Also, we can check the form of the electromagnetic functions for the case where we have an electrically charged source, which are written as:
\begin{equation}
    L(r)=-\frac{18 a^2 m}{5 r^5}, \quad L_F(r)=\frac{2 q_e^2 r}{3 a^2 m}, \quad F(r)=-\frac{9 a^4 m^2}{2 q_e^2 r^6}.
\end{equation}

In terms of the scalar $F$, the electromagnetic Lagrangian is
\begin{equation}
    L(F)=-\frac{2\ 2^{5/6} \sqrt[3]{3}  q_e^{5/3}}{5 a^{4/3} m^{2/3}}|F|^{5/6}.
\end{equation}
Once again, the result is consistent with those already existing in the literature \cite{Alencar:2024yvh}. This shows us that the method is effective in obtaining the source of some types of BBs in GR.

It is important to emphasize that in some cases it is not possible to represent the source of solutions using a scalar field that is purely canonical or purely phantom, thus requiring the consideration of partially phantom scalar fields \cite{Bronnikov:2022bud,Bolokhov:2024sdy,Crispim:2024lzf,Rodrigues:2025plw,GRAVMOD4}. This type of model demands a greater degree of freedom in the scalar field, thereby introducing additional functions to be determined.

\section{Holonomy corrected Schwarzschild BH}\label{S:HCS}
The Simpson-Visser case was made as a way to test the validity of the method. Now let us apply it to the Holonomy corrected Schwarzschild spacetime, and see if this proposal can be interpreted as a solution to the Einstein equations.

The Holonomy corrected Schwarzschild BH is described by the line element \eqref{lineel} with \cite{Alonso-Bardaji:2021yls}
\begin{equation}
g(r)=\left(1-\frac{2m}{r}\right), \qquad f(r)=\left(1-\frac{2m}{r}\right)\left(1-\frac{r_0}{r}\right),\label{coef_HCS}
\end{equation}
where $0<r_0<2m$ represents the radius of the throat of a WH with minimal area $4\pi r_0^2$ \cite{Moreira:2023cxy}.

Despite the metric coefficient being different, the causal structure of the Simpson-Visser solution is very similar to Holonomy corrected Schwarzschild BH. Therefore, it is possible that this spacetime, like the Simpson-Visser solution, can also be found from GR with the appropriate sources. Motivatingly, the geometry of this spacetime was investigated with regard to gravitational lensing effects, as well as the deflection of light into the strong and weak field regimes \cite{Soares1}. Subsequently, these analyses were extended to a system where the geometry is topologically charged due to the presence of a global monopole \cite{Soares2,Ahmed1}. Gravitational effects of this geometry have also been explored in the study of source field stability as well as usual and long-lived quasinormal modes \cite {MODOS1,MODOS2}, as well as in the context of tidal effects for a regularized geometric version \cite{Crispim:2025cql}.

Substituting the metric coefficients \eqref{coef_HCS} in \eqref{eqGR1}-\eqref{eqphi}, we find
\begin{eqnarray}
    \phi(r)&=&\sqrt{2} \tan ^{-1}\left(\frac{\sqrt{r-r_0}}{\sqrt{r_0}}\right),\qquad
    V(r)=\frac{r_0 (3 m+2 r)}{12 r^4},\label{V-HCS}\\
    L(r)&=&\frac{r_0 (9 m+4 r)}{6 r^4}, \quad
    L_F(r)=\frac{r_0 (3 m+r)}{4 q_m^2}, \quad F(r)=\frac{2q_m^2}{r^4}.\label{LF-HCS}
\end{eqnarray}
The electromagnetic functions, \eqref{LF-HCS}, obey the consistency condition \eqref{EMbound}.

\begin{figure}
    \centering
    \includegraphics[width=0.5\linewidth]{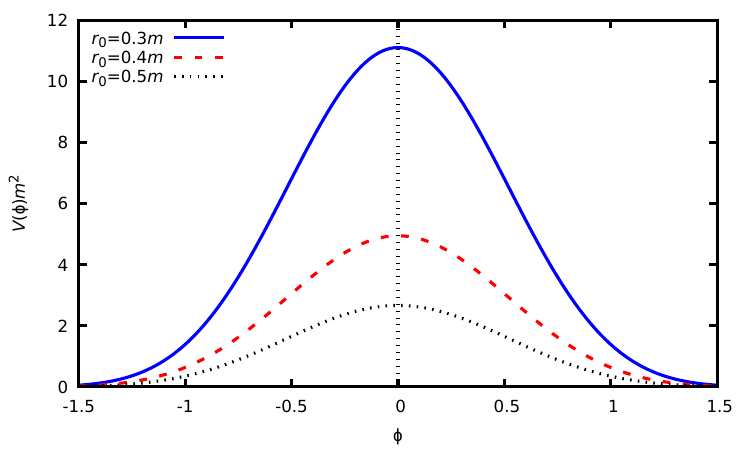}
    \caption{Behavior o the scalar field potential as a function of $\phi$ considering different values of $r_0$.}
    \label{fig:Vphi_LQG}
\end{figure}

We can still write the electromagnetic Lagrangian and the potential associated with the scalar field as
\begin{equation}
    L(F)=\frac{r_0 \left(4 \sqrt[4]{2} F^{3/4} \sqrt{q_m}+9 F m\right)}{12 q_m^2},\qquad V(\phi)=\frac{\cos ^6\left(\frac{\phi }{\sqrt{2}}\right) \left(3 m \cos \left(\sqrt{2} \phi
   \right)+3 m+4 r_0\right)}{24 r_0^3}.\label{L-V-HCS}
\end{equation}
In the limit of weak fields, $F\approx 0$, the dominant term is $F^{3/4}$, which is nonlinear. Expanding $V(\phi)$ to $\phi \approx 0$, we find
\begin{equation}
    V(\phi) \approx \frac{6 m+4 r_0}{24 r_0^3} - \frac{\phi ^2 (2 m+r_0)}{4 r_0^3}.
\end{equation}
Thus, we see that at zero the potential tends to a constant, and the first correction term is proportional to $\phi^2$, similar to the correction of a massive scalar field potential, similar to the Simpson-Visser case but with a different mass. In Fig. \ref{fig:Vphi_LQG}, we see the behavior of the potential, which resembles a potential barrier that grows as the value of $r_0$ decreases.

We found that, despite originally appearing in a different context, the Holonomy corrected Schwarzschild BH can be obtained through Einstein equations when coupled with NED and a phantom scalar field.

Just like we did in the Simpson-Visser case, let us also check how the source looks when considering an electrically charged solution. In this case, the scalar field remains the same, and the electromagnetic functions are written as:
\begin{equation}
    L(r)=-\frac{r_0 (9 m+2 r)}{6 r^4}, \quad L_F(r)=\frac{4 q_e^2}{3 m r_0+r r_0}, \quad F(r)=-\frac{r_0^2 (3 m+r)^2}{8 q_e^2 r^4}.\label{L-Eletric_HCS}
\end{equation}
The Lagrangian $L(r)$ is similar but not identical to the magnetic case.

It is possible to invert $F(r)$ in order to obtain the Lagrangian $L(F)$. However, the analytical expression is extensive and not very clear, so it is not worth writing about. It is important to note that for weak fields, the Lagrangian behaves as $F^{3/2}$, which is different from the magnetic case.

If we want a more compact analytical expression for the electromagnetic Lagrangian, we can use the auxiliary field $P^{\mu\nu}$, and thus we obtain the following:
\begin{equation}
    L(P)=\frac{r_0 \left(9 m P-2 \sqrt[4]{2} (-P)^{3/4} \sqrt{q_e}\right)}{12 q_e^2}.
\end{equation}
The Lagrangian form for the electric case in terms of $P$ is very similar to the form in terms of the scalar $F$ for the magnetic case, particularly regarding its asymptotic behavior.

\section{Polymerized BH} \label{S:POLY}
Another spacetime inspired by LQG is the polymerized BH. This spacetime is regular and, like the previous cases, has a throat hidden by an event horizon, thus being a model of a BB. The line element that describes this spacetime is written as \cite{Peltola:2008pa}:
\begin{equation}
    ds^2=\left(\sqrt{1-\frac{k^2}{r^2}}-\frac{2m}{r}\right)dt^2-\left(\sqrt{1-\frac{k^2}{r^2}}-\frac{2m}{r}\right)^{-1}\left(1-\frac{k^2}{r^2}\right)^{-1}dr^2-r^2\left(d\theta^2+\sin^2\theta d\varphi^2\right).\label{poly}
\end{equation}
The constant $k>0$ is the polymerized parameter and represents the radius of the throat with a minimal area $4\pi k^2$. It is interesting that there is also an event horizon, independent of the value of $k$, different from the Simpson-Visser spacetime. The horizon is located at $r_h=\sqrt{k^2+4m^2}$, and the radius of the throat is consistently smaller than the radius of the event horizon in this coordinate system. In addition to the original article \cite{Peltola:2008pa}, this polymerized solution was explored in the analysis of gravitational waves and nearly circular spiral orbits \cite{Yang:2024cnd} as well as in the study of gravitational lensing effects and light deflection in the strong and weak field regimes \cite{KumarWalia:2022ddq}.
If we try to map this spacetime as a solution of the equations, magnetically charged, the source can be described as a phantom scalar field with a NED, described by the functions:
\begin{eqnarray}
    \phi(r)&=&\cot ^{-1}\left(\frac{k}{\sqrt{r^2-k^2}}\right), \qquad V(r)= \frac{2 k^2 }{15 r^5}\left(6 m-\frac{r \sqrt{1-\frac{k^2}{r^2}} \left(3
   k^4+4 k^2 r^2+8 r^4\right)}{k^4}\right),\\
   L(r)&=&\frac{2 \left(6 k^2 m+5 r^3\right)}{5 r^5}+\frac{2 \sqrt{1-\frac{k^2}{r^2}} \left(r^2-k^2\right) \left(9 k^2+16
   r^2\right)}{15 k^2 r^4},\\
   L_F(r)&=&-\frac{\sqrt{1-\frac{k^2}{r^2}} \left(3 k^2+2 r^2\right)}{4 q_m^2}+\frac{3 k^2 m+r^3}{2 q_m^2 r},\quad F(r)=\frac{2 q_m^2}{r^4}.
\end{eqnarray}

Inverting $F(r)$, we can write the Lagrangian in terms of the electromagnetic scalar as:
\begin{equation}
    L(F)=\frac{3\ 2^{3/4}
   F^{5/4} k^2 m}{5 q_m^{5/2}}+\frac{\sqrt{2} \sqrt{F}}{q_m}+\sqrt{2-\frac{\sqrt{2} \sqrt{F} k^2}{q_m}}\frac{ \left(-9 \sqrt{2} F
   k^4-14 \sqrt{F} k^2 q_m+32 \sqrt{2} q_m^2\right)}{30 k^2 q_m^2}.
\end{equation}
In the week field limit, we find $L(F)\approx 32/15 k^2-F k^2/2 q_m^2$.

If we now calculate the sources of the solution considering the presence of an electric charge instead of a magnetic charge, we obtain the following:
\begin{eqnarray}
    L(r)&=&\frac{\sqrt{1-\frac{k^2}{r^2}} \left(27 k^4+16 k^2 r^2+32 r^4\right)}{15 k^2
   r^4}-\frac{18 k^2 m}{5 r^5}, \quad L_F(r)=\frac{4 q_e^2 r^2 \sqrt{1-\frac{k^2}{r^2}}}{3 k^4+2 r \sqrt{1-\frac{k^2}{r^2}} \left(3
   k^2 m+r^3\right)-k^2 r^2-2 r^4},\\
   F(r)&=&-\frac{\left(-3 k^4-2 r \sqrt{1-\frac{k^2}{r^2}} \left(3 k^2 m+r^3\right)+k^2 r^2+2
   r^4\right)^2}{8 q_e^2 r^8 \left(1-\frac{k^2}{r^2}\right)}.
\end{eqnarray}
These functions are more complicated than in the previous cases, so now we are no longer able to invert $F(r)$ analytically. However, we can still use the auxiliary field $P$ to write the Lagrangian $L(P)$, which is given by:
\begin{equation}
    L(P)=\frac{9 k^2 P \left(2\ 2^{3/4} m \sqrt[4]{-P}-\sqrt{q_e} \sqrt{4-\frac{2 \sqrt{2} k^2
   \sqrt{-P}}{q_e}}\right)}{20 q_e^{5/2}}+\frac{8 \sqrt{2-\frac{\sqrt{2} k^2 \sqrt{-P}}{q_e}}
   \left(k^2 \sqrt{-P}+2 \sqrt{2} q_e\right)}{15 k^2 q_e}.
\end{equation}
Thus, we have shown that we can consistently reconstruct LQG inspired BHs as solutions of GR with a phantom scalar field and NED, where this electrodynamics can be with either an electric or magnetic source.

\begin{figure}
    \centering
    \includegraphics[scale=0.7]{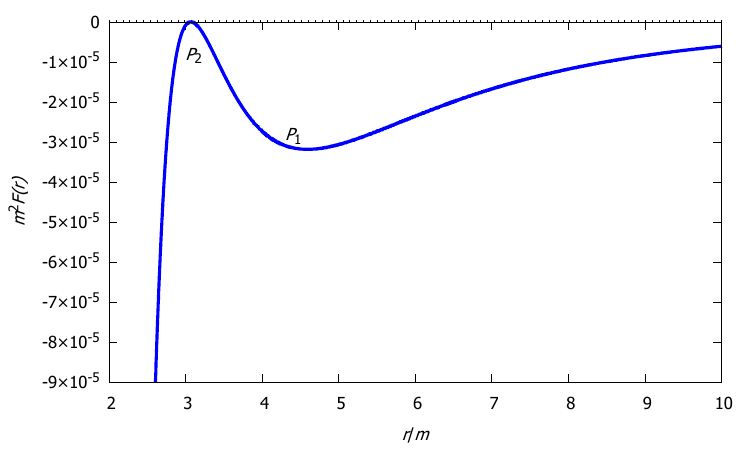}
    \includegraphics[scale=0.7]{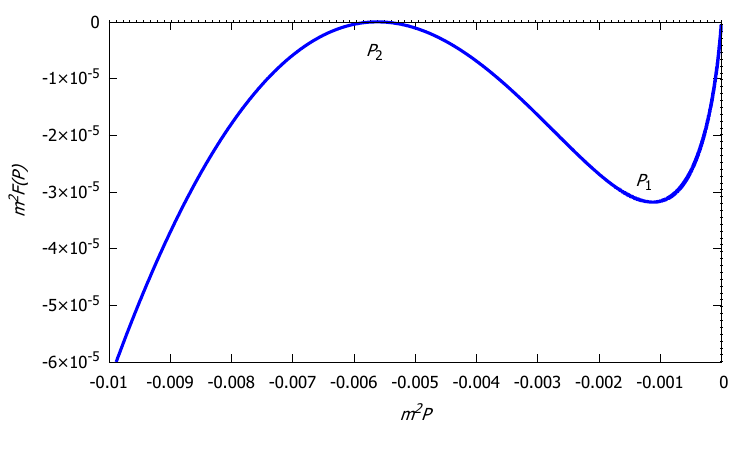}
    \includegraphics[scale=0.7]{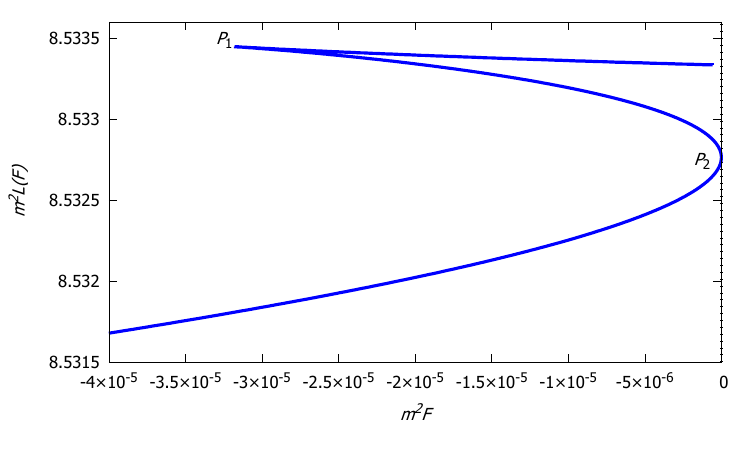}
    \caption{Plot of the functions $F(r)$, $F(P)$, and $L(F)$ to the polymerized solution \eqref{poly}, with $q_e=k=0.5m$. The extreme points $P_1$ and $P_2$ inform us about the points where the electromagnetic Lagrangian changes its behavior.}
    \label{fig:Eletro_functions}
\end{figure}

Despite not being able to analyze $L(F)$ analytically, we can still see the behavior of the electromagnetic functions graphically in Fig. \ref{fig:Eletro_functions}. We see why it is not possible to invert the function $F(r)$, since it has maxima and minima, making it impossible to invert analytically. The fact that the dependence $F(P)$ has extreme points reveals to us that the electromagnetic Lagrangian will exhibit different behaviors at different points in the spacetime.

Now that we know what types of sources generate these solutions in GR, we can use it to obtain properties about them.

\section{Energy Conditions}\label{S:EnerCond}
To determine whether the material content of a solution is usual or exotic, we must analyze the energy conditions associated with the stress-energy tensor. The energy conditions we will consider are the null (NEC), weak (WEC), dominant (DEC), and strong (SEC) energy conditions. Typically, models of BBs violate all energy conditions associated with the total stress-energy tensor in at least some regions of spacetime because they violate the NEC. However, here we will consider the conditions for each field separately.
The standard energy conditions are written as follows:
\begin{eqnarray}
&&NEC_{1,2}=WEC_{1,2}=SEC_{1,2}
\Longleftrightarrow \rho+p_{1,2}\geq 0,\label{Econd1} \\
&&SEC_3 \Longleftrightarrow\rho+p_1+2p_2\geq 0,\label{Econd2}\\
&&DEC_{1,2} \Longrightarrow \rho-p_{1,2}\geq 0,\label{Econd3}\\
&&DEC_3=WEC_3 \Longleftrightarrow\rho\geq 0.\label{Econd4}
\end{eqnarray}
Subscript 1 refers to the case where the energy density is taken together with the radial pressure, while subscript 2 corresponds to the combination with the tangential pressure. Subscript 3, in turn, is used when only the energy density is considered, as in $WEC_3$, or when it is taken together with both pressures, as in $SEC_3$.

The fluid quantities are related to the components of the stress-energy tensor as
\begin{eqnarray}
T^{\mu}{}_{\nu}={\rm diag}\left[\rho,-p_1,-p_2,-p_2\right]\, \quad \mbox{if} \quad f(r)>0,\label{EMT}\\
T^{\mu}{}_{\nu}={\rm diag}\left[-p_1,\rho,-p_2,-p_2\right]\, \quad \mbox{if} \quad f(r)<0.\label{EMTOut}
\end{eqnarray}
Here, $\rho$ is the energy density, $p_1$ is the radial pressure, and $p_2$ is the tangential pressure.

The fluid quantities for the electromagnetic field do not change inside or outside an event horizon. These are then written as
\begin{equation}
    \rho[E]=\frac{L}{2}=-p_1[E],\qquad p_2[E]= -\frac{L}{2}+\frac{2q^2L_F}{r^4}.
\end{equation}
For the scalar field, the expressions depend on the presence of the event horizon. In this case, the fluid quantities are given by
\begin{eqnarray}
    \rho[\phi]=V-f \phi'^2=-p_2[\phi],\qquad p_1[\phi]= -V-f\phi'^2,
    \quad \mbox{if} \quad f(r)>0,\\
    \rho[\phi]=V+f\phi'^2,\qquad p_1[\phi]= -V+f\phi'^2=p_2[\phi].\quad \mbox{if} \quad f(r)<0.
\end{eqnarray}

The energy conditions for the Simpson-Visser case have already been analyzed for each field in the literature \cite{Rodrigues:2023vtm}. Therefore, we now need to check for solutions inspired by LQG.

It is important to note that NEC is also included in the other conditions. This reveals to us that if NEC is violated, the others will also be violated. However, if the null energy condition is satisfied, this does not guarantee that the others will also be satisfied.
\subsection{Holonomy corrected}
Considering the functions related to the electromagnetic field and the scalar field, obtained in the previous section, we find the expressions to the energy conditions for the Holonomy corrected Schwarzschild BH. For the electromagnetic sector, the expressions are written as
\begin{eqnarray}
&&NEC_{1}[E] \Longleftrightarrow 0, \quad NEC_{2}[E]
\Longleftrightarrow \frac{r_0 (3 m+r)}{2 r^4}\geq 0, \quad
SEC_3[E] \Longleftrightarrow \frac{r_0 (9 m+2 r)}{6 r^4}\geq 0,\label{Econd1EMSol1}\\
&&DEC_{1}[E]=2WEC_{3}[E] \Longrightarrow \frac{r_0 (9 m+4 r)}{6 r^4}\geq 0, \qquad DEC_{2}[E] \Longrightarrow \frac{r_0}{6 r^3}\geq 0.\label{Econd2EMSol1}
\end{eqnarray}
This reveals that NED does not violate the energy conditions and this is valid for the entire spacetime.

For the scalar field, in regions where $f(r)>0$, the energy conditions are
\begin{eqnarray}
&&NEC_{1}[\phi] \Longleftrightarrow -\frac{r_0}{r^3} \left(1-\frac{2m}{r}\right)\geq 0, \qquad NEC_{2}[\phi]
\Longleftrightarrow  0,\label{Econd1phi+sol1} \\
&&DEC_{1}[\phi]=-SEC_{3}[\phi] \Longrightarrow \frac{r_0 (3 m+2 r)}{6 r^4}\geq 0, \quad DEC_{2}[\phi]=2WEC_3[\phi] \Longrightarrow \frac{r_0 (15 m-4 r)}{6 r^4}\geq 0,\label{Econd2phi+sol1}
\end{eqnarray}
and for regions where $f(r)<0$, we have
\begin{eqnarray}
&&NEC_{1}[\phi]=NEC_{2}[\phi]\Longleftrightarrow\frac{r_0}{r^3} \left(1-\frac{2m}{r}\right)\geq 0,\label{Econd1phi-sol1} \\
&&SEC_3[\phi] \Longleftrightarrow\frac{r_0 (10 r-27 m)}{6 r^4}\geq 0, \quad WEC_3[\phi] \Longleftrightarrow\frac{r_0 (8 r-9 m)}{12 r^4}\geq 0,\label{Econd2phi-}\\
&&DEC_{1}[\phi]=DEC_{2}[\phi] \Longrightarrow \frac{r_0 (3 m+2 r)}{6 r^4}\geq 0.\label{Econd3phi-sol1}
\end{eqnarray}
Since at least one of the inequalities of NEC is always violated, all energy conditions will have at least one of their inequalities violated for the phantom scalar field.

Therefore, even if the electromagnetic field does not violate the energy conditions, the phantom scalar field will violate them.

\subsection{Polymerized solution}
Let us now analyze the energy conditions for the polymerized solution case. We will start with the electromagnetic part, as it is valid for the entire spacetime. These are given by:
\begin{eqnarray}
&&NEC_{1}[E] \Longleftrightarrow 0, \quad NEC_{2}[E]
\Longleftrightarrow \frac{6 k^2 m+2 r^3}{2 r^5}-\sqrt{1-\frac{k^2}{r^2}}\frac{ \left(3 k^2 +2 r^2\right)}{2 r^4}\geq 0,\label{Econd1EMsol2}\\ 
&&SEC_3[E] \Longleftrightarrow \frac{18 k^2 m}{5 r^5}-\frac{\sqrt{1-\frac{k^2}{r^2}} \left(27 k^4+16 k^2 r^2+32 r^4\right)}{15 k^2 r^4}\geq 0,\label{Econd2EMsol2}\\
&&DEC_{1}[E]=2WEC_{3}[E] \Longrightarrow \frac{2 \left(6 k^2 m+5 r^3\right)}{5 r^5}+\frac{2 \sqrt{1-\frac{k^2}{r^2}} \left(-9 k^4-7 k^2 r^2+16 r^4\right)}{15
   k^2 r^4}\geq 0,\label{Econd3EMsol2}\\
&&DEC_{2}[E] \Longrightarrow -\frac{3 k^2 m}{5 r^5}+\frac{\sqrt{1-\frac{k^2}{r^2}} \left(9 k^4+2 k^2 r^2+64 r^4\right)}{30 k^2
   r^4}+\frac{1}{r^2}\geq 0.\label{Econd4EMsol2}
\end{eqnarray}

For the scalar field, in regions where $f(r)>0$, the energy conditions are
\begin{eqnarray}
&&NEC_{1}[\phi] \Longleftrightarrow -\frac{2k^2}{r^4} \left( \sqrt{1-\frac{k^2}{r^2}}-\frac{2 m}{r}\right)\geq 0, \qquad NEC_{2}[\phi]
\Longleftrightarrow  0,\label{Econd1phi+sol2} \\
&&DEC_{1}[\phi]=-SEC_{3}[\phi] \Longrightarrow \frac{8 k^2 m}{5 r^5}+\frac{4 \sqrt{1-\frac{k^2}{r^2}} \left(-3 k^4 r-4 k^2 r^3-8 r^5\right)}{15 k^2 r^5}\geq 0,\label{Econd2phi+sol2} \\
&&DEC_{2}[\phi]=2WEC_3[\phi] \Longrightarrow \frac{28 k^2 m}{5 r^5}-\frac{2 \sqrt{1-\frac{k^2}{r^2}} \left(21 k^4+8 k^2 r^2+16 r^4\right)}{15 k^2 r^4}\geq 0,\label{Econd3phi+sol2}
\end{eqnarray}
and for regions where $f(r)<0$, we have
\begin{eqnarray}
&&NEC_{1}[\phi]=NEC_{2}[\phi]\Longleftrightarrow\frac{2k^2}{r^4} \left( \sqrt{1-\frac{k^2}{r^2}}-\frac{2 m}{r}\right)\geq 0,\label{Econd1phi-sol2} \\
&&SEC_3[\phi] \Longleftrightarrow\frac{8 \sqrt{1-\frac{k^2}{r^2}} \left(9 k^4+2 k^2 r^2+4 r^4\right)}{15 k^2 r^4}-\frac{48 k^2 m}{5 r^5}\geq 0,\label{Econd2phi-sol2} \\
&&WEC_3[\phi] \Longleftrightarrow \frac{\sqrt{1-\frac{k^2}{r^2}} \left(9 k^4-8 k^2 r^2-16 r^4\right)}{15 k^2 r^4}-\frac{6 k^2 m}{5 r^5}\geq 0,\label{Econd3phi-sol2}\\
&&DEC_{1}[\phi]=DEC_{2}[\phi] \Longrightarrow \frac{8 k^2 m}{5 r^5}-\frac{4 \sqrt{1-\frac{k^2}{r^2}} \left(3 k^4+4 k^2 r^2+8 r^4\right)}{15 k^2 r^4}\geq 0.\label{Econd4phi-sol12}
\end{eqnarray}
For the case of the scalar field, it is obvious that the $NEC_1[\phi]$ inequality is always violated in such a way that the NEC is never satisfied. This result is expected since the scalar field is phantom-like.

For the case of the electromagnetic field, it is not very clear whether the NEC is satisfied or not. However, if we expand the inequality $NEC_2[E]$ for points far from the BH, we obtain $NEC_2[E]\Longleftrightarrow-\frac{k^2}{r^4}\geq 0$, which is clearly not satisfied. This shows us that there are points outside the event horizon where the NEC will be violated.

\section{Conclusion}\label{S:conclusion}

In this work, we show that spherically symmetric and static BB spacetimes inspired by LQG can be interpreted as solutions of GR. For this, we constructed the formalism considering the coupling between the gravitational theory with NED and a non-free scalar field. The equations of motion were written in the most general form possible for the chosen line element, and we considered both magnetically charged and electrically charged cases.

The combination of the scalar field and NED is necessary since, for general BB cases, the components of the Einstein tensor do not exhibit the symmetries ${G^0}_{0} = {G^1}_{1}$ or ${G^0}_{0} = {G^2}_{2}$, which means that the fields individually cannot satisfy the Einstein equations. In general, the form of the functions related to the scalar field does not depend on whether the charge is electric or magnetic. Thus, changing the choice of the source only modifies the functions $L$, $L_F$, and $F$, while $\phi$ and $V(\phi)$ remain unchanged, regardless of the source. For all cases, we consider $\epsilon = -1$, ensuring that the scalar field is a phantom field for all solutions. This requirement is necessary to guarantee that the scalar field remains real.

The first model we studied was the Simpson-Visser spacetime. This model was chosen because it has been extensively studied in the literature and served as a consistency test for our method. The results obtained, both for magnetic and electric charges, differ from those presented in \cite{Rodrigues:2023vtm,Alencar:2024yvh}. However, this difference arises because of the chosen coordinate system, so despite being different, our results are consistent with those found in the literature.

As the second model, we chose the Schwarzschild solution with holonomic corrections. The causal structure of this solution is quite similar to that in the case of RBH in the Simpson-Visser model. However, unlike Simpson-Visser, we do not have traversable WH cases in both directions, since $0<r_0<2m$, while in the Simpson-Visser case, the parameter $a$ can take values greater than $2m$. 

The functions related to the source fields, in terms of the radial coordinate, for the magnetically charged case, are given by equations \eqref{V-HCS} and \eqref{LF-HCS}. We can also write $L(F)$ and $V(\phi)$ for this model, as in equation \eqref{L-V-HCS}. From the form of expression \eqref{L-V-HCS}, we note that, in the weak field limit ($F \approx 0$), the dominant term in the electromagnetic Lagrangian is not linear but rather $F^{3/4}$. 
For the electrically charged case, the functions related to the electromagnetic sector are given by \eqref{L-Eletric_HCS}. The form of the function $L(r)$ is quite similar to that in the magnetic case. However, the analytical form of $L(F)$ is quite complicated, which is why we work with the auxiliary field $P$. In the weak field limit, the dominant term is $P^{3/4}$, similar to what happens in the magnetic case.

For the third case, the functions related to the source fields are much more complex than in the previous cases. For example, the function $F(r)$ is sufficiently complicated that it cannot be inverted to obtain $r(F)$. The reason for this becomes more evident in Fig. \ref{fig:Eletro_functions}, where we observe the presence of two extrema in $F(r)$. This feature results in two changes in the behavior of $L(F)$, making the function $L(F)$ multivalued.

Once the fields have been separated, we analyzed the energy conditions for each field individually. Since the Simpson-Visser case has been extensively studied in the literature, we did not address it here. For the second model, we found that all the energy conditions related to the electromagnetic field are always satisfied, whereas the scalar field can violate all the energy conditions, both inside and outside the event horizon, which is consistent with a phantom scalar field. For the third model, some of the analytical expressions are not very clear. However, we observe that, for points far from the event horizon, the inequality related to $NEC[E]$ is not satisfied, ensuring that, at least in some regions, the NEC is violated. Moreover, for the scalar field, it becomes more evident that the NEC is violated in all regions of spacetime. In summary, NEC violations are consistent with the behavior expected for BB solutions, such as those discussed here \cite{Zaslavskii:2010qz}.

The stability properties of solutions are known to depend critically on their source characteristics \cite{Gonzalez:2008wd,Bronnikov:2013coa,Bronnikov:2021xao}. Consequently, classifying the source types capable of producing regular solutions is essential, which is the principal aim of this preliminary investigation. A natural continuation of our work is to assess viability through dynamical and observational probes, such as quasinormal modes, potential echoes, and long-lived modes in ultracompact regimes, in order to compare the stability of regular BHs and BB families sourced by different matter sectors. In particular, we will confront the sources obtained here with anisotropic fluids \cite{Duran-Cabaces:2025sly,Franzin:2023slm}.

It is well known that photons do not follow null geodesics in NED. As a consequence, photons may propagate with speeds larger or smaller than the vacuum speed of light. Causality in NED has been extensively investigated in the context of standard regular BHs and is also linked to the energy conditions \cite{dePaula:2024yzy,Tomizawa:2023vir}. In future work, we intend to extend this causality analysis, and its connection to the energy conditions, to BB spacetimes supported by NED, as discussed here. We will also compare the resulting phenomenology in the effective and background metrics, focusing on light rings, shadows, and lensing \cite{dePaula:2023ozi}. Taken together with shadow and lensing constraints, these tests will help delineate which sources support solutions that are regular, causal, and observationally consistent.

It is also possible to consider these spacetimes within alternative theories of gravity, such as $f(R)$ gravity or other modified gravity models, so that the matter sources are modified. This can mitigate violations of the energy conditions and reduce the need for phantom fields \cite{GRAVMOD4,GRAVMOD3}.

\section*{Acknowledgments}
The authors would like to thank Conselho Nacional de Desenvolvimento Cient\'{i}fico e Tecnol\'ogico (CNPq) and Funda\c c\~ao Cearense de Apoio ao Desenvolvimento Cient\'ifico e Tecnol\'ogico (FUNCAP) for the financial support.

\bibliographystyle{apsrev4-1}
\bibliography{ref.bib}
\end{document}